\documentclass[12pt]{article}
\usepackage[pctex32]{graphics}
\usepackage[pctex32]{graphics}
\begin{document}
~
\vspace{1cm}
\begin{center} {\Large \bf  Instability of Tachyon Supertube in Type IIA  G\"odel Spacetime}
                                                  
\vspace{2cm}

                      Wung-Hong Huang\\
                       Department of Physics\\
                       National Cheng Kung University\\
                       Tainan, Taiwan\\

\end{center}
\vspace{2cm}
\begin{center} {\large \bf  Abstract} \end{center}
We study the tachyon supertube probes  in a type IIA supergravity background which is a stringy-like G\"odel spacetime and contains closed timelike curve.   In the case of small value of $f$, which is a parameter of the background, we use the Minahan-Zwiebach tachyon action to obtain a single regular tube solution and argue that the tube is a BPS D2-brane.   However, we find that the fluctuation around the tube configuration has a negative-energy mode.   This means that the tachyon supertube, despite being a BPS configuration, develops an instability in the pathological spacetime with closed timelike curve  which violates the causality.    

\vspace{2cm}
\begin{flushleft}
E-mail:  whhwung@mail.ncku.edu.tw\\
\end{flushleft}
\newpage
\section {Introduction}
G\"odel had found a homogeneous and simply connected universe with closed timelike curves (CTCs) [1]. This particular solution is not an isolated pathology and in fact the CTCs are a generic feature of gravitational theories, in particular in higher dimensions with or without the supersymmetry [2].    CTCs are not linked to the presence of strong gravitational fields.  They are global in nature and will violates the causality.  Physicists usually label the spacetimes with CTCs as pathological, unphysical  and forget about them.   However, Hawking had formulated the chronology protection conjecture, with a quantum mechanism enforcing it by superselecting the causality violating field configurations from the quantum mechanical phase space [3].

   In a previous paper [4] we have calculated the energy-momentum tensor of a scalar field propagating in a one-parameter family of solutions that includes
the four-dimensional generalized G\"odel spacetime [5]. As the parameter is varied, we had shown that the energy-momentum tensor becomes divergent precisely at the onset of CTCs.  This gives a support to the Hawking's chronology protection conjecture.  In [6]  Gauntlett et. al. calculated the holographic energy-momentum tensor in a deformation of $AdS_5\times S^5$ to investigate the problem in the string theory.  Their results, however, showed that the holographic energy-momentum tensor remains finite even when the CTCs appear.\footnote {The inconsistence between [4] and [6] has not yet been clarified.} 

  In recent many physicists hope that the string theory could provide a mechanism to rule out the pathological solutions or eliminate the CTCs [6-13].  For example, Herdeiro showed that the causality bound corresponds to a unitarity bound in the CFT [7], and therefore that the over rotating spacetimes, which will form CTCs, are not genuine solutions of string theory.  Boyda et. al. found that holography can serve as a chronology protection agency [8].  Astefanesei et. al. argued that the  conjectured AdS/CFT correspondence may teach us something about the physics in spacetimes containing closed timelike curve [9].  Caldarelli et. al. showed that a violation of the Pauli exclusion principle in the phase space of the fermions is thus intimately related to causality violation in the dual geometries [10]. 

  The philosophy behind this is that, in string theory there are dynamical extended objects which may provide us with probes suited particularly well to study non-local issues like closed timelike curves. 

   One of the interesting objects in the string theory is the supertube found by Mateos and Townsend [14].   It is a tubular bound state of D0-branes, fundamental strings (F1) and D2-branes, which is supported against collapse by the angular momentum generated by the Born-Infeld (BI) electric and magnetic fields.  Drukker et. al. studied the dynamics of cylindrical D2-brane in Type IIA G\"odel universe and claimed that the supertube  develops an instability in the CTC region despite being a BPS configuration [11].  

   The investigations in [11] use the supertubes as probes, in which the supertubes  are described as BPS D2-branes in DBI action.   As well-known that, in the spirit of the Sen's conjecture [15], the BPS branes could also be viewed as tachyon kinks of non-BPS branes in higher dimension, it is therefore interesting to adopt the tachyon tube [16, 17] as a probe to investigate the problem.    This is the work of this paper.   

   In section II, we first describe  a type II supergravity background which behaves as the G\"odel Spacetime [18] and contains closed timelike curve.  Then we use the Minahan-Zwiebach tachyon action [19] to find the single regular tube solution with circular cross section in the stringy-like G\"odel spacetime.   We will see that the energy of the single tubular configuration comes entirely from the D0 and strings at critical Born-Infeld (BI) electric field.  Thus the solution is supersymmetric [14] and the tachyon supertube solution is a BPS configuration.   In section III we calculate the fluctuation spectrum around the tube solution and show that there is a negative-energy mode.   This means that the tachyon  supertube, despite being a BPS configuration, develops an instability in the pathological spacetime with closed timelike curve which violates the causality.   Our result provides an alternative proof of Drukker's result [11] while using the tachyon field theory. We make a conclusion in the last section.

\section {Tube Solution in a Stringy-like G\"odel Spacetime}
We adopt a specific solution of type IIA supergravity with rotation in a single plane.  This solution is a stringy-like G\"odel  spacetime preserving one quarter of the maximal number of supersymmetries [18].   It  has a nontrivial metric in three dimensions, which we parameterize by the time coordinate $t$, and polar coordinates $r$ and $\theta$ in the plane. There is also a NS-NS flux as well as RR 2-form and 4-form fluxes.   The flux fields are used to render the spacetime supersymmetric. The metric and fluxes of this stringy-like G\"odel spacetime  are given by 
$$ds^2 = - \left[dt + fr^2d\theta\right]^2 + dz^2 + dr^2 + r^2 d\theta^2 + \delta_{ij}dx^i dx^j,  \eqno{(2.1)}$$
$$B_{NS} = f r^2dz\wedge d\theta \ , \ \ \ \ C_3 = fr^2d\theta\wedge dt \wedge dz \ ,  \ \ \ \ C_1 = - fr^2d\theta \ .  \eqno{(2.2)}$$
The spacetime has CTC when $r > f^{-1}$ [18].  However, as the stringy-like G\"odel universe is homogeneous there are CTCs through every point in space.  

  In this section we will adopt the MZ tachyon action [19] to analyze the supertube in the stringy-like G\"odel spacetime (2.1).  The Minahan-Zwiebach (MZ) tachyon action is a derivative truncation of the BSFT action of the non-BPS branes [20], which embodies the tachyon dynamics for unstable D-branes in (super)string theories and was first proposed as a simplified action to capture the desirable properties of string theories.    The action had been successfully used to study the phenomena of kink condensation and vortex condensation in the unstable non BPS branes [21].  The results support the Sen's conjecture of the `Brane Descent Relations' of tachyon condensation.  It had also been used to investigate the problems of the recombination of intersecting branes [22].   

  The Minahan-Zwiebach tachyon action of  the non-BPS D3 brane, including the Wess-Zumino terms, is described by [18,19]
$$ S = -{\cal T}_3 \int V(T) \left(1+ (\partial_\mu T)^2 + {1\over 4} {\cal F}_{\mu\nu}^2\right) + \int V(T) ~ dT \wedge  ( C_3 + C_1 \wedge {\cal F})\,. \eqno{(2.3)}$$
The field strength ${\cal F}$ includes the BI gauge field strength $F_{BI}$ that are turned on in the brane and those induced by the NS field  strength $B_{NS}$ in (2.2), i.e. 
$${\cal F} = F_{BI} + B_{NS}, ~~~~ F_{BI}= E \, dt\wedge dz + B \, dz\wedge d\theta . \eqno{(2.4)}$$
in which we allow for a time-independent electric field $E$ and magnetic field $B$.   The supertube we considered is a cylindrical D2-brane which is extended in the $z$ direction as well as the angular direction $\theta$ at fixed radius $r$ about the origin.   It is supported against collapse by the angular momentum generated by the Born-Infeld (BI) electric field $E$ and magnetic fields $B$ [14].   Under these conditions the Lagrangian is
$$ {\cal L} = - 2\pi V(T) \,r \left[1 + T'^2 + {1\over2} E^2 (f^2 R^2-1) - f E(B - f r^2) + {(B - f r^2)^2\over 2r^2} - f r(1-  E)T'\right], \eqno{(2.5)}$$
in which the tachyon field is a function of radius as we consider only the case of circular tube.   The associated field equation is 
\\
$$2V(T) \left[T''(x) + {T'\over r} - f (1-E)  \right] - V'(T)\left[1- T'^2 +  f r (1-E)T' + {1\over2} E^2 (f^2 R^2-1)\right.$$
$$\left.  - f E(B - f r^2) + {(B - f r^2)^2\over 2r^2} \right] = 0, \eqno{(2.6)}$$
\\
which, however, could not be solved exactly and we shall adopt some approximations.

   In the case of $f\ll 1$ we have found the solution
$$T_c(r) \approx {~B\over {\sqrt 2}~}\, ln(r/r_0) + {f (1- E_c)\over 4}\, r^2, \eqno{(2.7)}$$
in which the critical value of electric field $E_c$ is defined by 
$$E_c = {\sqrt 2} - \left(1 + {\sqrt 2\over 2}\right) f B .\eqno{(2.8)}$$
The above solution is consistent with our previous paper [17] in the case of flat space, i.e. $f =0$.   The value of  $r_0$ in above is an arbitrary integration constant.   As the value of $|T(r)_c|$ becomes zero near $r=r_0$ the radius of tachyon tube will depend on the value of  $r_0$.   The tube radius and $r_0$ are determined by the BI EM field, i.e., the charges of D0 and strings on the brane.  It is noted that, as that in the previous cases [17], the above tube solution is irrelevant to the function form of the tachyon potential $V(T)$. 

   To proceed we define the electric displacement defined by $\Pi = \partial {\cal L} /\partial E$ and thus from (2.5) 
$$\Pi =   - 2\pi V(T) \left[E (f^2 R^2-1) r - f (B - f r^2) r  - T' f B r^2\right], \eqno{(2.9)}$$
The associated energy density defined by $H = \Pi E - L$  becomes
$$H  =  2\pi V(T) \left[\left(1 + T'^2 - {1\over2} E^2 (f^2 R^2-1) + {(B - f r^2)^2\over 2r^2}\right) r - f r^2T' \right]. \eqno{(2.10)}$$
In figure 1 we plot the typical behaviors of function $H(r)$ which shows that there is a peak at finite radius for the solution (2.7).
\\
\\
\scalebox{1}{\hspace{4cm}\includegraphics{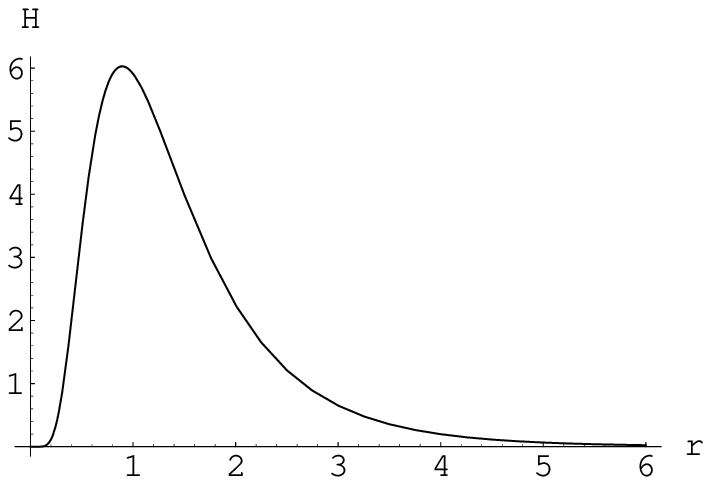}}\\ 
\hspace{2cm}{\it ~~~Figure 1. The behaviors $H(r)$ in (2.10) in the case of $B =2$ , $f = 0.01$ and $V(T) = e^{- T^2}$.  There is a peak at finite radius r which specifies the size of the  circular cross-section tube of solution (2.7).}\\
\hspace{1cm}

We can now use the regular tachyon solution (2.7) to evaluate the D0 charge $q_0$ and F-string charge $q_1$, which are defined by [16,17]
$$ q_1 = {1\over 2\pi}\int d \varphi \, \Pi\, ,~~~~~~ q_0 = {1\over 2\pi}\int d \varphi \,  B\,.\eqno{(2.11)}$$ 
respectively.   It is seen that energy density ${\cal U}$ could be expressed as 
$${\cal U} = E_c \,q_1 +  \,q_0. .\eqno{(2.12)}$$
As the energy density of tachyon tube  is just given by the sum of charges it carries this solution represents a BPS tube, as dictated by supersymmetry [14,16,17]. 

  It is also interesting to mention the physical meaning of the critical value of electric field $E_c$.  Substituting the tachyon solution (2.7) into the Lagrangian we see that 
$${\cal L_c} \approx - B \left[{B\over r} - {2-{\sqrt 2}\over 2} f\, r + {1-{\sqrt 2}\over 2}B f \right] V(T_c).\eqno{(2.13)}$$
Thus increasing $E$ to its `critical' value $E = E_c$ would reduce the D2-brane  tension to zero if the magnetic field were zero [14].  This implies that the tachyon tube has no energy associated to the tubular D2-brane tension; its energy comes entirely from the electric and magnetic fields, which can be interpreted as `dissolved' strings and D0-branes, respectively.  The energy from the D2-brane tension has been canceled by the binding energy released as the strings and D0-branes are dissolved by the D2-brane.  The phenomena that the tubular D2-brane tension has been canceled was crucial to have a supersymmetric tube configuration [14]. 

   We conclude this section a comment. The critical electric in DBI action is $E_c=1$ [14] which is different from that in the MA tachyon action.  The inconsistence may be traced to the fact that the MZ tachyon action is just a derivative truncation of the BSFT action of the  non-BPS branes [17].   However, we hope that the truncated action could capture desirable properties of the brane theories.

\section{Fluctuation and Instability of Tachyon Supertube}
Let us now consider the fluctuation $t$ around the tubular solution 
$$T(r) = T(r)_c + t(r),  \eqno{(3.1)}$$
Substituting the tubular solution $T(r)_c$ in (2.7) into the action (2.5) and considering only the quadratic terms of fluctuation field $t(r)$ we obtain
$$S = - 2\pi\int dr\left[\left(1+({\sqrt 2} - f r)\right)~r~\hat{t'}^2 + \left({B\over 8 r}\left (4 B -f~ r (4({\sqrt 2}-1) + \right.\right.\right.~\hspace{3cm}~$$
$$\left.\left.\left.\left.~~~~({\sqrt 2}-2)(4 + B^2)r\right)\right) Ln(r/r_0) + 2B\left(({\sqrt 2}-2)f r^2 -B(1+ fr - {\sqrt 2}fr)\right)Ln(r/r_0)^2~ \right)\hat{t}^2 \right]$$
$$ =   - 2\pi\int dw\left[\hat t \,{\partial^2 \hat t\over\partial w^2} +{2B^2- B^4 w^2\over 4} ~ \hat{t}^2 + \delta H ~ \hat{t}^2\right ],~~\hspace{5cm}~\eqno{(3.2)}$$
\\
in which 
$$\delta H = f{e^w\over 8}\left[4({\sqrt 2}-1)B^2 +2(-2 +{\sqrt 2})B^2 w^2 \right.e^w~\hspace{4cm}~ $$
$$~\hspace{3cm}~\left.+ 4 w\left((1-{\sqrt 2})(B^2+B^4) - ({\sqrt 2}-2)(4+B^3)e^w \right)\right],\eqno{(3.3)}$$
and we have used the partial integration, field redefinition 
$$\hat{t} \equiv V(T_c)^{-1\over2} \,t,~~~~~~~~~with ~~~~V(T_c) = e^{-T_c^2}, \eqno{(3.4)}$$
and new variable 
$$w= ln\left({r/ r_0}\right) - ln\left[1+ (\sqrt2 -1)f{r/ r_0}\right].\eqno{(3.5)}$$
Without the $\delta H$ term ( i.e. $f=0$ and space becomes flat) we see that the fluctuation $\hat t$ obeys a Schr\"odinger equation of a harmonic oscillator, thus the mass squared for the fluctuation is equally spaced and specified by an integer $n$, 
$$m^2_t=  2n\,B^2 , ~~~~~ \ n \geq 0 .\eqno{(3.6)}$$
Thus there is no tachyonic fluctuation, the mass tower starts from a massless state $\Psi(w)_0$ and has the equal spacing. This result is consistent with the identification of the tachyon tube as a tubular BPS D2-brane.  This is the result obtained in [17]. 

 Now we can use the perturbation method of quantum mechanism  to perform the calculation including the $\delta H$ term effect.   We then find that the energy of the ground state $\Psi(w)_0$  becomes
$$\delta E_{\Psi(w)_0}= 2\pi \int dw\, \delta H \, |\Psi(w)_0|^2 = {\pi e^{1\over4B}\over 2}\left [(1-\sqrt{2})B(1-B)^2 + e^{3\over4B}(2-\sqrt{2})(2-B)\right]$$
$$\approx -\,{(\sqrt 2 -1)\pi\over 4}f\, |B|^3,~~~~~~~ if ~~|B| \gg 1,\eqno{(3.7)}$$
which is a negative value. (Note that to have CTCs the parameter $f $ must be positive.)  We thus conclude that the tachyon supertube, despite being a BPS configuration, develops an instability in the pathological spacetime with closed timelike curve.
\section{Discussions}
In normal background of spacetime  a BPS state satisfies a Bogolmonyi 
bound, which implies that the energy is minimized for a given charge, and 
state is absolutely stable.  However, in the pathological spacetime with 
closed timelike curve, a BPS state will be unstable, as shown in [11] using DBI action and in this paper by using the tachyonic BDI action.

   In fact, the instability of the BPS supertube in type IIA  G\"odel spacetime could be easily seen from the analyses of DBI action, as briefly described in below.  The world-volume theory on the supertube is just that of a D2-brane in curved background, which includes the Dirac-Born-Infeld and Wess-Zumino terms
$$S = - \int e^{-\phi}\sqrt{-\det( {\cal G} + {\cal F})} -\int( C_3 + C_1 \wedge {\cal F})\,,\eqno{(4.1)}$$
where $\cal G$ is the pullback of the metric and $C_1$ and $C_3$ the pullbacks of the RR potentials. The field strength ${\cal F}$ includes the BI gauge field strength $F_{BI}$ that are turned on in the brane and those induced by the NS field  strength $B_{NS}$ in (2.2).   Under these conditions the Lagrangian  of a static straight tube of circular cross section with radius $R$ becomes [11]
$${\cal{L}} = - \sqrt{ R^2 + \Delta^{-1} \tilde{B}^2  - \tilde{E}^2 {R^2}\Delta } 
- f R^2  + f R^2 E \, ,\eqno{(4.2)}$$
where
$$ \Delta \equiv  1 -f^2 R^2 \ , \ \ \tilde{B} \equiv  B - f R^2  \ , \ \ \tilde{E} \equiv  E - { f\tilde{B}}\Delta^{-1} \,.\eqno{(4.3)} $$
The momentum conjugate to $E$ takes the form
$$\Pi \equiv {\partial{\cal L}\over \partial E}  = {\tilde{E} R^2 \Delta \over \sqrt{ R^2 + \Delta^{-1} \tilde{B}^2  - \tilde{E}^2 {R^2}\Delta }} + f R^2.   \eqno{(4.4)}$$ 
Solving the above equation we have the relation
$$\tilde{E} = s ~ {\tilde{\Pi}\over R \Delta}\sqrt{ R^2  \Delta^2 +  \tilde{B}^2\over R^2  \Delta^2 +  \tilde{\Pi}^2}, ~~~~~ \tilde{\Pi} = \Pi - f R^2 , ~~~~~s = sign  ({\tilde{\Pi}/ \Delta}).  \eqno{(4.5)}$$ 
\\
The corresponding Hamiltonian density per unit length is
$$ {\cal H}(R) \equiv \Pi E - {\cal L}  = {1\over R \Delta}\sqrt{ R^2  \Delta^2 +  \tilde{B}^2\over R^2  \Delta^2 +  \tilde{\Pi}^2} ~ \left[R^2  \Delta^2 + s ~ \tilde{\Pi}^2\right] + {f\over \Delta} \tilde{B} \tilde{\Pi} + f R^2.   \eqno{(4.6)}$$
\\
Then, the tubular bound state of  the F-string and D0-brane may be formed if the energy  ${\cal H}(R)$ has a minimum at finite radius $R$.   For example in figure 1 we plot the radius-dependent energy  for the case of $f = 0.1, ~ B =\Pi = 5$.  The tube with radius $R = 5$ is found. 
\\
\\
\scalebox{1}{\hspace{4cm}\includegraphics{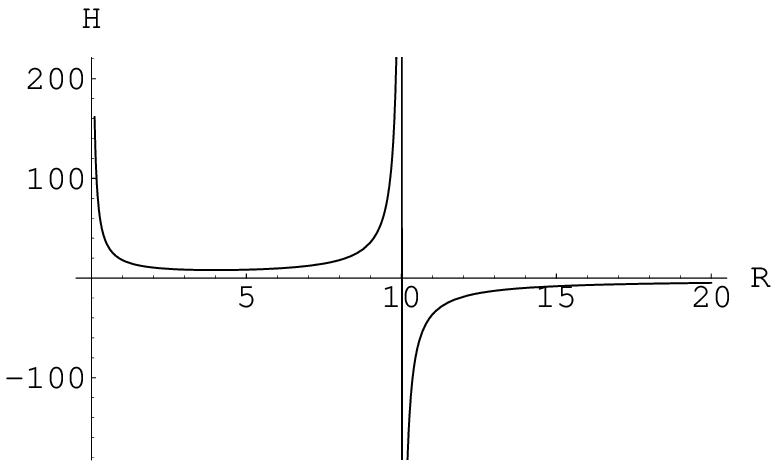}}\\ 
\hspace{2cm}{\it ~~~Figure 2: The energy density for the case of $f = 0.1, ~ B =\Pi = 5$.  The energy density has a local minimum at $R_c = B = 5$  which is the radius of the tube.  As the energy density of the tube solution is just given by the sum of charges,  it represents a BPS tube solution [14].  However, the energy density at $R= f^{-1}$ become $ - \infty $, the BPS tube in the stringy G\"odel universe is in fact unstable.} 
\hspace{1cm}
\\
\\
The functional form of energy ${\cal H}(R)$ in (4.6) can be analyzed if $B = \Pi$.  In this case we have a simple relation 
$${\cal H}(R) = {B^2 + R^2 - 2 f B R^2 \over R - f R^2}, \eqno{(4.7)}$$
and it has a solution tube with radius $R_c = B$.   The associated energy density is
$$H(R_c) = 2 B \sim q_0 + q_1 . \eqno{(4.8)}$$
As the tube energy density is just given by the sum of charges it carries (Note that in the case of $B = \Pi$ the charges is proportional to $B$), this solution represents a BPS tube, as dictated by supersymmetry [14].   However, from figure  2  we can see that the energy ${\cal H}(R= f^{-1}) \rightarrow - \infty $, which signals that the BPS tube in the stringy G\"odel universe may be unstable. 

   In conclusion, in this paper we have investigated the properties of tachyon supertubes in a type IIA supergravity background which behaves as a stringy-like G\"odel Spacetime and contains closed timelike curve.   In the case of small value of $f$, which is a parameter of the background, we use the Minahan-Zwiebach tachyon action to obtain a single regular tube solution and argue that the tube is a BPS D2-brane.  We have investigated the fluctuations around the tachyon supertube configuration and found that there is a negative-energy mode.   This means that the tachyon supertube, despite being a BPS configuration, develops an instability in the pathological spacetime with closed timelike curve  which violates the causality.   This provides an alternative proof of Drukker's result [11] while using the tachyon field theory.    

   The dynamics of (extended) probes in spacetimes containing CTCs has been studied extensively in recent [6-13].   It becomes clear that in general the worldvolume theories develop instabilities (negative energy modes).  Our investigations suggest a possible way to use the tachyon brane to probe the pathological spacetime with closed timelike curve.   Using the tachyon brane as a probe to study other pathological spacetimes with closed timelike curve is an interesting problem.  It remains to be studied.

\newpage
\begin{enumerate}
\item  K.~G\"odel, ``An example of a new type of cosmological solutions of Einstein's field equations of gravitation,'' Rev.\ Mod.\ Phys. 21 (1949) 447.
\item R.~C.~Myers and M.~J.~Perry,``Black holes in higher dimensional spacetimes,'' Annals Phys.172 (1986) 304; G.~W.~Gibbons and C.~A.~R.~Herdeiro, ``Supersymmetric rotating black holes and causality violation,'' Class.\ Quant.\ Grav.\ 16 (1999) 3619 [hep-th/9906098].
\item S.~W.~Hawking, ``The Chronology protection conjecture,'' Phys.\ Rev.\ D  46 (1992) 603;  M.~J.~Cassidy and S.~W.~Hawking, ``Models for chronology selection,'' Phys.\ Rev.\ D 57 (1998) 2372 [hep-th/9709066].
\item  Wung-Hong ~Huang,``Chronology Protection In Generalized G\"odel Spacetime,'' Phys.\ Rev.\ D  60 (1999) 067505 [hep-th/0209091].
\item M.~J.~Reboucas and J.~Tiomno, ``On The Homogeneity Of Riemannian Spacetimes Of G\"odel Type,'' Phys.\ Rev.\ D  28 (1983) 1251.
\item  Jerome P. Gauntlett, Jan B. Gutowski and Nemani V. Suryanarayana ,``A deformation of $AdS_5\times S^5$,'' Class.\ Quant.\ Grav.\  21 (2004) 5021 [hep-th/0406188].
\item  C.~A.~R.~Herdeiro, ``Special properties of five dimensional BPS rotating black holes,'' Nucl.\ Phys.\ B 582 (2000) 363 [hep-th/0003063].
\item E.~K.~Boyda, S.~Ganguli, P.~Horava and U.~Varadarajan, ``Holographic protection of chronology in universes of the Goedel type,'' Phys.\ Rev.\ D 67 (2003) 106003 [hep-th/0212087].
\item D. Astefanesei, R. B. Mann, E. Radu, ``Nut Charged Space-times and Closed Timelike Curves on the Boundary,'' [hep-th/0407110].
\item M. M.~Caldarelli, D. Klemm  and P. J. Silva, ``Chronology Protection in anti-de~Sitter,'' [hep-th/0411203].
\item N.~Drukker, B.~Fiol and J.~Simon, ``Godel's universe in a supertube shroud,'' Phys.\ Rev.\ Lett.\  91 (2003) 231601 [hep-th/0306057]; N.~Drukker,
``Supertube domain-walls and elimination of closed time-like curves in string
theory,'' [hep-th/0404239]; D.~Brace, ``Closed geodesics on Goedel-type backgrounds,'' JHEP  0312 (2003)  021 [hep-th/0308098].
\item Y.~Hikida and S.~J.~Rey, ``Can branes travel beyond CTC horizon in Goedel universe?,'' Nucl.\ Phys.\ B  669 (2003) 57 (2003) [hep-th/0306148]; E.~G.~Gimon and P.~Horava, ``Over-rotating black holes, Goedel holography and the hypertube,'' [hep-th/0405019]; D.~Brace, ``Over-rotating supertube ackgrounds,''
[hep-th/0310186].
\item  L.~Jarv and C.~V.~Johnson,``Rotating black holes, closed time-like curves, thermodynamics, and the enhancon mechanism,'' Phys.\ Rev.\ D  67 (2003) 066003 [hep-th/0211097]; D.~Israel, ``Quantization of heterotic strings in a Goedel/anti de Sitter spacetime and chronology protection,'' JHEP  0401 (2004)042 [hep-th/0310158]. 
\item D. Mateos and P. K. Townsend, ``Supertubes'', Phys. Rev. Lett. 87 (2001) 011602 [hep-th/0103030]; R. Emparan, D. Mateos and P. K. Townsend, ``Supergravity Supertubes'', JHEP 0107 (2001) 011 [hep-th/0106012]; D.~Mateos, S.~Ng and P.~K.~Townsend, ``Tachyons, supertubes and brane/anti-brane systems'', JHEP  0203 (2002) 016 [hep-th/0112054]; D. Bak, K. M. Lee, ``Noncommutative Supersymmetric Tubes'',  Phys. Lett. B509 (2001) 168 [hep-th/0103148]; D. Bak and S. W. Kim, ``Junction of Supersymmetric Tubes,'' Nucl. Phys.  B622 (2002) 95 [hep-th/0108207]; D. Bak and A. Karch, ``Supersymmetric Brane-Antibrane Configurations,'' Nucl. Phys.  B626 (2002) 165 [hep-th/011039]; D. Bak and N. Ohta, ``Supersymmetric D2-anti-D2 String,'' Phys. Lett.  B527 (2002) 131;  Wung-Hong Huang, ``Condensation of Tubular D2-branes in Magnetic Field Background'', Phys. Rev. D70 (2004) 107901 [hep-th/0405192];  Wung-Hong Huang, ``Condensation Tubular Solutions in NS5-brane, Dp-brane and Macroscopic Strings Backgrounds'', JHEP  0502 (2005) 061 [hep-th/0502023].
\item A. Sen, ``Tachyon Condensation on the Brane Antibrane System'', JHEP 9808 (1998) 012, [hep-th/9805170]; ``Descent Relations Among Bosonic D-branes'',  Int.\ J.\ Mod.\ Phys. A14 (1999)  4061, [hep-th/9902105]; ``Supersymmetric world-volume action for non-BPS D-branes,''  JHEP  9910 (1999) 008 [hep-th/9909062];  ``Tachyon Dynamics in Open String Theory'', [hep-th/0410103].
\item L. Martucci and P. J. Silva, "Kinky D-branes and straight strings of open string tachyon effective theory",   JHEP 0308 (2000)  026 [hep-th/0306295];
C. Kim, Y. Kim, O-K. Kwon, and P. Yi, ``Tachyon Tube and Supertube,''  JHEP 0309 (2003) 042 [hep-th/0307184].
\item Wung-Hong Huang ``Tachyon Tube on non-BPS D-branes,''  JHEP 0408 (2004) 060 [hep-th/0407081]; Wung-Hong Huang ``Entropy and Quantum States of Tachyon Supertube,''  JHEP 0412 (2004) 002 [hep-th/0410264].
\item T.~Harmark and T.~Takayanagi, ``Supersymmetric G\"odel universes in string theory,'' Nucl.Phys. B662 (2003) 3 [hep-th/0301206];  H.~Takayanagi,
``Boundary states for supertubes in flat spacetime and G\"odel universe,''
JHEP  0312 (2003) 011 [hep-th/0309135].
\item  J.  A. Minahan and B.  Zwiebach, ``Effective Tachyon Dynamics in Superstring Theory'',  JHEP  0103 (2001) 038 [hep-th/0009246]; J.  A.  Minahan and B.  Zwiebach, ``Field Theory Models for Tachyon and Gauge Field String Dynamics'',   JHEP 0009 (2000)  029 [hep-th/0008231].
\item M. R. Garousi, `` Tachyon couplings on non-BPS D-branes and Dirac-Born-Infeld action,''  Nucl.Phys. B584 (2000) 284 [hep-th/0003122]; A. Sen, ``Dirac-Born-Infeld Action on the Tachyon Kink and Vortex,''  Phys. Rev. D68 (2003) 066008 [hep-th/0303057];  E.A. Bergshoeff, M. de Roo, T.C. de Wit, E. Eyras, S. Panda, `` T-duality and Actions for Non-BPS D-branes,'' JHEP 0005 (2000) 009 [hep-th/003221].
\item K. Hashimoto and S. Nagaoka,``Realization of Brane Descent Relations 
in Effective Theories'', Phys. Rev. D66 (2002) 0206001  [hep-th/0202079].
\item  Wung-Hong Huang, ``Recombination of Intersecting D-branes in Tachyon Field Theory'' , Phys.Lett. B564 (2003) 155 [hep-th/0304171].
\end{enumerate}
\end{document}